# SecureTime:
# Secure Multicast Time Synchronization

Robert Annessi, Joachim Fabini, and Tanja Zseby

Institute of Telecommunications, TU Wien, Austria
{robert.annessi, joachim.fabini, tanja.zseby}@tuwien.ac.at

**Abstract.** Due to the increasing dependency of critical infrastructure on synchronized clocks, network time synchronization protocols have become an attractive target for attackers. We identify data origin authentication as the key security objective and suggest to employ recently proposed high-performance digital signature schemes (Ed25519 and MQQ-SIG)) as foundation of a novel set of security measures to secure multicast time synchronization. We conduct experiments to verify the computational and communication efficiency for using these signatures in the standard time synchronization protocols NTP and PTP. We propose additional security measures to prevent replay attacks and to mitigate delay attacks. Our proposed solutions cover 1-step mode for NTP and PTP and we extend our security measures specifically to 2-step mode (PTP) and show that they have no impact on time synchronization's precision.

## 1  Introduction

Time synchronization protocols have become an essential building block of various applications. The deployment of time synchronization for controlling system clocks of critical infrastructures, such as telecommunication, industrial automation, avionics, or energy distribution raises awareness with respect to the security of time synchronization protocols. Attacks on time synchronization can originate faulty sensor reports, endanger control decisions, and adversely affect the overall functionality of applications depending on it. For this reason, time synchronization protocols have to be secured whenever they are used outside of fully protected network environments.

In this paper, we abstract the multicast time synchronization scenario from any specific implementation such as the Network Time Protocol (NTP) or the Precision Time Protocol (PTP). In multicast communication a sender periodically sends time synchronization messages that can be received by multiple receivers. In contrast to unicast communication, time synchronization messages are not sent to receivers individually in multicast communication. Instead, the sender sends each time synchronization message only once, and the message is replicated by the communication network along distinct paths whenever needed. In this way, multicast communication is more efficient when handling multiple receivers, because the load on the sender is reduced and the number of copies that traverse a network is minimized. In NTP [1], broadcast[1] communication is primarily used for more efficient communication because the increasing use of network time synchronization results in additional load on time senders and networks. In PTP [2], on the other hand, multicast communication is inherently integrated into the protocol.

---

[1] Broadcast communication can be seen as a special case of multicast communication where messages are sent to the entire group of receivers.

While unicast communication can be appropriately secured, the two most widely used time synchronization protocols, NTP and PTP, do not protect multicast time synchronization messages adequately, leaving applications vulnerable to attacks. Since no general solution exists as of now, securing multicast time synchronization remains a challenging open problem that requires significant attention [3]. In this paper, our aim is to maintain both security as well as the lowest possible degradation of time synchronization's accuracy.

**Contributions**

1. We identify data origin authentication as the key security objective in multicast time synchronization and suggest recently proposed high-performance signature schemes (Ed25519 and MQQ-SIG)) to secure multicast time synchronization against substitution and impersonation attacks.
2. We propose an additional set of security measures including sequence numbers and session keys to prevent replay attacks, as well as a novel method to mitigate delay attacks, which consists of a secure delay measurement procedure, dynamic offset correction limitation, and a receiver-specific delay measurement interval based on one-way network delays and clock drift. We furthermore extend the security measures to 2-step mode in PTP, which additionally entails random numbers and a hash function to link SYNC and FOLLOWUP messages so that security is provided without requiring the SYNC message to be signed.
3. We conduct a security analysis, which shows that the proposed security measures entirely prevent substitution attacks, pre-play attacks, and replay attacks, while delay attacks are mitigated. For delay attacks, we give upper bounds on the delays that can be introduced maliciously by an adversary.
4. We conducted performance measurements and show that using the proposed high-speed signature has only low impact on the precision in 1-step mode (28 µs or 75 µs) and, furthermore, introduce only very low communication overhead (32 B or 64 B plus the size of the sequence number) per message. In 2-step mode, our security measures have practically no impact on time synchronization's precision.

## 2  Background

Due to the lack of access control in many communication networks, cryptographic schemes are required to ensure that receivers can verify that messages have been, indeed, sent by the claimed sender. This security property is called data origin authentication. Data origin authentication directly affects integrity, authenticity, and non-repudiation. In the context of time synchronization, confidentiality is generally of less concern.

## 2.1 Data Origin Authentication

Digital signatures provide data origin authentication but the main downside of today's digital signature schemes, such as RSA, DSA, and ECDSA, is that they involve high computational cost and substantial penalty in terms of delay, both in the sender and in the receiver. Consequently, it is widely believed that digital signatures are roughly 2 to 3 magnitudes slower than Message Authentication Codes (MACs) [4], such that signing each packet is not a practical solution. In particular, the time that is needed for signing packets and verifying digital signatures is detrimental to the precision of time synchronization protocols. We revise this assumption as we investigate the potential of recently proposed high-performance digital signature schemes as basis for data origin authentication in multicast time synchronization.

## 2.2 Time Synchronization

There are basically two distinct phases in time synchronization protocols: time offset correction and network delay measurement. Time offset correction is conducted at regular intervals while network delay measurements are conducted irregularly. The specific intervals depend on the time synchronization protocol and on the system configuration. Delay measurements are conducted over a unicast connection while time offset corrections are multicast in PTP and optionally broadcast in NTP.

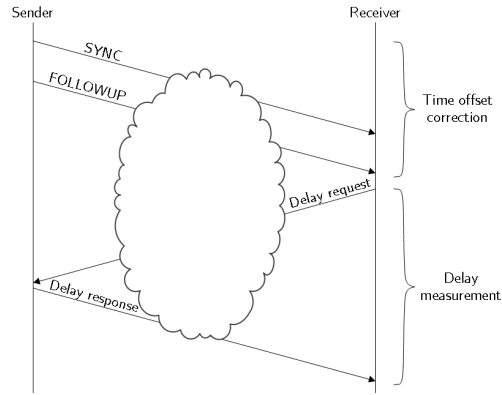

**Fig. 1.** Time offset correction and delay measurement.

Time offset corrections can either be done in a single message (1-step mode) supported by both NTP and PTP or with two messages (2-step mode) supported by PTP. In 2-step mode, time synchronization is split into two messages, a SYNC and a FOLLOWUP message. The SYNC message is just used as a marker; the FOLLOWUP message contains the exact point in time when the SYNC message was sent. Figure 1 depicts the (2-step) time offset correction and delay measurement.

Traversing network components between sender and receiver introduces variable delays. PTP allows network devices to improve the precision of time synchronization by measuring and communicating to receivers the delay they have introduced to SYNC messages (in the FOLLOWUP message). Such network devices that actively support the time synchronization protocol are called transparent clocks. Communicating the delay that transparent clocks introduced to a particular message requires these devices to modify the content of the FOLLOWUP message, which poses a serious challenge in the light of data origin authentication. We solve this challenge by introducing security measures specific to the 2-step mode so that transparent clocks are supported.

# 3 Threat Model

Henceforth, we assume that receivers already know the (long-term) public key of the particular sender and can trust that this public key is both valid and correctly bound to the sender. I.e., we assume that the certified public key has been transmitted initially in a reliable and authenticated manner.

## 3.1 Network Model

The overall purpose of time synchronization protocols is to convey time information in order to compensate for clock variations. We consider communications involving one sender and a potentially large number of receivers. Messages are delivered from the sender to each receiver through an unreliable, potentially lossy communication network, such as the Internet. Furthermore, we assume that neither network devices nor receivers can be trusted, since the larger the number of devices the higher the probability that at least one is compromised. Besides that, receivers and network devices might not even be controlled by the same entity that controls the sender. The network only forwards packets, it does not provide any security guarantee such as confidentiality, integrity, and authenticity so that messages may be read, modified, dropped, or injected by entities other than the intended receivers. This corresponds to the Dolev-Yao threat model [5].

## 3.2 Adversary Model

We assume that the adversary does not permanently prevent the communication between sender and receiver; i.e., the receiver will receive most messages. The adversary has full control over the network and can selectively capture, drop, resend, inject, delay, and alter messages arbitrarily with negligible delay. The computational power of the adversary is limited but not necessarily bound to that of the sender or the receivers; i.e., the adversary can use more powerful devices and larger storage. Furthermore, the adversary can compromise an arbitrary number of receivers and learn all their secrets. An important aspect of assuming a powerful adversary is that if security properties hold against such powerful adversary, they also hold against less capable adversaries.

## 3.3 Attacks

The adversary aims to make receivers adhere to false time values or to degrade the accuracy of the time synchronization. For this purpose, the adversary can conduct various attacks [6]:

*Substitution Attack* In a substitution attack [7], the adversary intercepts valid time synchronization messages during transmission and modifies them in such a way that receivers accept the forged messages as if they had been sent by the original sender. In this way, false time information can be distributed to receivers.

*Replay Attack* An adversary can record time synchronization messages and replay them without modification at a later time, since successful verification of a message does not certify the correctness of the message's send time [4]. In this way, inaccurate information can be intentionally provided to receivers.

*Pre-play Attack* In contrast to replay attacks, the adversary injects messages at an earlier point in time in pre-play attacks, before the sender has even sent the message. For this purpose, the adversary needs to be able to predict future messages.

*Removal Attack* A removal attack results when time synchronization messages are intentionally dropped by the adversary. As previously stated, we assume that the adversary does not drop most messages because this would then be an issue of availability rather than authenticity or integrity.

*Delay Attack* To conduct a message delay attack, an adversary intercepts time synchronization messages and delays these messages artificially for some time before forwarding them. If the effective delay is calculated maliciously, the clock of the intended receiver(s) can be manipulated [6]. Since the time synchronization protocol has no information about the underlying communication infrastructure, symmetric network delay between sender and receiver is assumed; i.e., the one-way delay from sender to receiver is the same as the delay from receiver to sender. Message delay attacks exploit this assumption of symmetric delay by maliciously introducing asymmetric delay in such a way that the receiver synchronizes to an inaccurate time.

*Flooding Attack* In Denial of Service (DoS) attacks [8], the computational or storage capacities of receivers are exhausted in order to prevent or delay the reception of messages. Such DoS attacks can be conducted, for example, by an adversary sending an excessive number of time synchronization messages to a receiver. Data origin authentication cannot prevent message flooding attacks but it can reduce their impact, as it can provide means for receivers to distinguish valid from invalid messages (up to some number of packets per second).

Data origin authentication can not or only partially prevent message flooding and message removal attacks. Those attacks only degrade precision, however. Henceforth, we will focus on preventing attacks that could make receivers synchronize to a false time and therefore could cause serious disturbance of applications that rely on accurate time: substitution attacks, replay attacks, pre-play attacks, and delay attacks.

## 4 Security Measures

In this section, we introduce a set of security measures to secure multicast time synchronization. The security these measures provide against the attacks identified in Section 3 will be analyzed in Section 5. We will use the notation in Table 1 on the next page.

**Table 1.** Notation

| | |
|---|---|
| $t_{arr}$ | The local arrival time of a message. |
| $t_{last}$ | The local arrival time of the last (valid) message seen by the receiver. |
| $\rho_{max}$ | The maximum drift of the receiver's clock relative to that of the sender. |
| $\delta_{max}$ | The maximum one-way network delay. |
| $\delta_{min}$ | The minimum one-way network delay. |
| $\epsilon_m$ | $2 \cdot (\delta_{max} - \delta_{min})$ |
| $\epsilon_1$ | $3 \cdot (\delta_{max} - \delta_{min})$ |
| $\epsilon_2$ | $4 \cdot (\delta_{max} - \delta_{min}) + 2 \cdot \delta_{max} \cdot \rho_{max}$ |

The specific values for $\delta_{min}$ and $\delta_{max}$ depend on the type of network (local, private network or public Internet) and on the quality of service measures in place. If the adversary is assumed to transfer messages with negligible delay, as done in Section 3, $\delta_{min}$ should be set to 0. If $\delta_{max}$ is chosen too small, false alarms may be triggered.

### 4.1 Data Origin Authentication

Data origin authentication is at the heart of the security measures we propose. The evaluation in [9] has revealed that most existing data origin authentication schemes suffer from severe limitations when securing multicast time synchronization. As a solution one could either improve one of the two schemes, Receiver driven Layered Hash-chaining (RLH) [10] or Time Valid Hash to Obtain Random Subsets (TV-HORS) [11], or, alternatively, try a different approach. We opted for the latter and propose a different data origin authentication scheme in which every single message is signed independently, although common perception is that such an approach is impractical due to the fact that conventional signature schemes are computationally too expensive. In recent years, however, novel signature schemes have been proposed that offer previously unrivaled performance. Employing such novel, high-performance signature schemes can mitigate the negative performance impact of traditional schemes. In Section 6, we evaluate the computational efficiency and the communication overhead of this approach. We would like to stress, however, that the additional security measures we propose do not depend on this specific data origin authentication scheme as any other scheme could be used as drop-in replacement as long as it provides existential unforgeability[2].

### 4.2 Freshness of Messages

Replay attacks can be prevented in the context of multicast time synchronization by ensuring that once a message has been received it is not accepted again at a later point in time; i.e., the freshness of the message is ensured. To this end, a data origin authentication scheme alone does not suffice because successful verification of a message does not imply that the message has never been received before. To ensure the freshness of messages, we suggest that the sender includes a sequence number in each message that is monotonically increasing; the sequence number starts with zero[3] for the first message and increases by one with every message. Receivers and the sender need to store only one sequence number as state information. For every received message (that has a valid signature) the receiver checks the freshness of the message by verifying that the included sequence number is greater than the locally stored one. If this is the case, the message is accepted and the local state updated accordingly; otherwise, the message is discarded.

---

[2] Informally, existential unforgeability means that the adversary can forge a signature to any message of his choice only with negligible probability.

This strict handling of sequence numbers allows for lost messages due to transmission errors but prevents a message from getting accepted by the receiver if that message was overtaken by another message during transmission. While this may sound like a disadvantage at the first glance, it is actually beneficial in the context of time synchronization. Messages that were overtaken in transit entail a larger network delay than other messages. Since network delay variation has a negative effect on the receiver's notion of time, is is beneficial to the precision of time synchronization when those delayed messages get discarded.

### 4.3 Session Keys

Since the sequence number is of fixed size, it will overflow eventually. A long-term attacker could capture all messages and start replaying them as soon as the sequence number overflows. To prevent such long-term attack, the sender signs a session public key (with its long-term secret key) as well as the current sequence number and sends both to the receiver at the start of the communication.

The receiver verifies the correctness with the sender's long-term public key. Before the sequence number overflows, the communication is restarted with a fresh session key-pair, and the sequence number is again reset to zero. This introduction of session keys, used for only a limited number of messages, provides not only security against replay attacks by long-term attackers but also reduces the time an adversary has to attack a specific key-pair. In this way, session keys also reduce the pressure on the employed data origin authentication scheme.

### 4.4 Dynamic Time Offset Correction Limitation

NTP limits both offset correction for individual messages as well as offset correction during a time span to global, fixed values. Those global values, however, only slow down but cannot bound delay attacks. In PTP, offset correction limitation does not exist at all so that time offsets can be arbitrary large; i.e., time can be set back or forth by years via a single message.

We, on the other hand, limit the maximum time offset correction that a receiver applies to $(t_{arr} - t_{last}) \cdot \rho_{max}$. This dynamic time offset correction limitation restricts the maximum influence that individual time synchronization messages can have on the receiver's notion of time. For each message (with a valid signature and a valid sequence number) the receiver limits the time offset correction to $(t_{arr} - t_{last}) \cdot \rho_{max}$.

---

[3] There is no need for a random initial sequence number for two reasons: (1) the sequence number is part of the time synchronization messages, which is cryptographically signed so that an adversary cannot modify it unnoticedly. And (2), an adversary cannot even gain information about the start of the communication from the sequence number since the adversary cannot know how often the sequence number has overflowed.

### 4.5 Receiver-Specific Delay Measurement Interval

In PTP, delay measurements happen at regular, non-specified intervals. In broadcast NTP, network delay is only measured once in the beginning. To ensure that the time offset between sender and receiver cannot be changed arbitrarily between two consecutive delay measurements, we use a receiver-specific delay measurement interval of $\frac{\epsilon_m}{\rho_{max}}$. As we will show in Section 5.4, this receiver-specific delay interval together with dynamic time offset correction limitation and secure delay measurement is sufficient to bound delay attacks.

### 4.6 Secure Delay Measurement Procedure

We propose a secure delay measurement procedure, in which a single delay measurement consists of two authenticated unicast messages: (1) a request from a receiver to the sender and (2) a response from the sender to the receiver. To ensure that the delay measurement is not artificially delayed, the receiver includes a timestamp in the delay request to the sender. Furthermore, the receiver waits at most $2 \cdot \delta_{max}$ for the sender's response. When the delay request arrives at the sender, the sender checks if the difference of the included timestamp and the sender's local clock is in the interval $[\delta_{min}, \delta_{max}]$. If it is not, the sender sends an error message to the receiver because of a potential delay attack[4]. Otherwise, the observed delay is so small that the sender cannot distinguish it from network delay variation. Therefore, the sender answers with a delay response message to the receiver that includes the timestamp of the sender. The receiver now conducts the same check (observed time difference within $[\delta_{min}, \delta_{max}]$). These checks ensure that the delay measurements themselves have not been artificially delayed more than the maximum network delay variation.

In the delay response, the sender furthermore includes the current sequence number as well as an identifier of its session's public key. The receiver updates its sequence number state to the number entailed in the delay response message and sets $t_{last}$ to the time of the delay measurement. Furthermore, the receiver checks that the session public key identifier matches the session public key (if it does not match, the client needs to restart the communication). In this way, the transition to a new session can be prevented maliciously at most for one delay measurement interval (plus the timeout) until the receiver notices. For this reason, an adversary cannot get more time to attack an old key.

## 5 Security Analysis

We assume that the adversary is in a privileged network position and conducts various, potentially severe attacks (as described in Section 3). We furthermore assume that the adversary does not know the secret key of the sender but he may know the sender's public key as he can compromise any number of receivers. With respect to the data origin authentication scheme, we assume that it provides existential unforgeability.

---

[4] At the very first delay measurement, the receiver may ignore the sender's error message and just set its local time to the timestamp in the error message, as sender and receiver may not yet be synchronized within a precision of $\leq \epsilon_m$.

In this section we argue that an adversary cannot make a receiver adhere to false time information by subsitution attacks, impersonation attacks, or replay attacks when the receiver receives time synchronization messages from a honest sender and both, receiver and sender, employ the previously outlined security measures. Furthermore, we will show that the maximum impact of delay attacks is very limited and can be bound.

### 5.1 Substitution attack

In a substitution attack, the adversary first intercepts messages including the corresponding authentication information. Eventually, the adversary substitutes parts of an intercepted message such that the receiver believes that the modified message originated from the sender.

We argue that a substitution attack can only be conducted successfully when breaking the data origin authentication scheme. In a substitution attack, we distinguish two cases: (1) The adversary modifies the message in such a way that the resulting message is identical to a message the adversary has intercepted. For the intercepted message, the adversary also intercepted the corresponding, valid signature. However, this case is equivalent to either a replay attack or to a delay attack, depending on whether or not the adversary dropped the original message. The security of the security measures against replay attacks as well as against delay attacks is analyzed separately later. (2) The adversary modifies the message in such a way that it is new. In order to make a receiver accept the new message, the adversary needs to provide valid authentication information (under the sender's public key).If the adversary can generate valid authentication information with non-negligible probability he can also efficiently forge signatures, which contradicts existential unforgeability provided by the data origin authentication scheme.

### 5.2 Pre-play Attack

In order to get a pre-played message accepted by the receiver, the adversary needs to generate a valid signature. Again, if the adversary can do that efficiently he can also break the existential unforgeability of the underlying data origin authentication scheme.

### 5.3 Replay Attack

In a replay attack, the adversary injects a message that was intercepted before and that included valid authentication information provided by the sender. We argue that a receiver that employs our security measures discards replayed messages unless the adversary breaks the underlying data origin authentication scheme. To this end, we distinguish two cases: (1) The adversary prevented the original message from reaching the receiver, which is equivalent to a delay attack where the adversary holds a message for some time and later forwards it to the receiver. Delay attacks, specifically, are analyzed later. (2) The adversary did not prevent the original message from reaching the receiver.

Since the receiver received the original message, the receiver has updated its sequence number state with the sequence number from the received message. If no other message reached the receiver in the meantime, the sequence number in the replayed message is identical to the receiver's local state. For this reason, the receiver will discard the replayed message.[5] When other messages have reached the receiver in the meantime, we need to distinguish two sub-cases: (2a) The maximum sequence number value was not reached. In this case, the sequence number in the replayed message is smaller than the receiver's local state, and the receiver will therefore discard the replayed message.[5] (2b) The maximum sequence number value was reached. In this case, the sequence number of the replayed message may actually be greater than the last seen sequence number stored locally by the receiver at that time because the sequence number was reset to zero when the new session was started. The authentication information of the replayed message, however, is not valid anymore since the sender has switched to a new key-pair for the new session. The adversary, therefore, needs to generate authentication information that is valid under the new key-pair in order to make the receiver accept the message. This is equivalent to an impersonation attack, however, which was analyzed previously.

### 5.4 Delay Attack

In a delay attack, the adversary intercepts a message, delays it for some time, and forwards it later to the receiver. The goal of the delay attack is to influence the receiver's notion of time maliciously. In contrast to the other attacks, message delay attacks cannot be prevented entirely. We will show, however, that our security measures can mitigate delay attacks by providing two upper bounds to the time offset between sender and receiver: (1) the maximum time offset so that a delay attack can go unnoticed $\epsilon_1$, and (2) the maximum time offset before the receiver does notice the delay attack $\epsilon_2$. Before proving the upper bounds $\epsilon_1$ and $\epsilon_2$ we first establish two building blocks: (1) that the maximum time offset between sender and receiver that is unnoticed at a delay measurement equals $\epsilon_m$ and (2) that the maximum time offset correction that can be applied between two consecutive delay measurements equals $\epsilon_m$ as well. Figure 2 on the facing page depicts the bounds and the building blocks.

Let's assume that the time offset between sender and receiver is $> \epsilon_m$ at a delay measurement interval. According to the security measures, the receiver sends a delay measurement request to the sender that includes the timestamp of the receiver. We distinguish two cases: (1a) The adversary drops the receiver's delay measurement request. After waiting for the server's response for the timeout period $2 \cdot \delta_{max}$, however, the receiver notices the delay attack. (1b) The adversary does not drop the receiver's delay measurement request so that it arrives at the sender. The sender compares the receiver's time entailed in the delay measurement request to the sender's local time and notices that the time difference is $> \epsilon_m$. For this reason, the sender will return an error message back to the receiver. The adversary can now either drop the sender's response, which is equally noticed by the receiver after the timeout, or the adversary does not drop the sender's error message and the receiver notices the attack from the error message. For this reason, the time offset at a delay measurement will be noticed when $> \epsilon_m$ and therefore must be at $\leq \epsilon_m$ to go unnoticed.

---

[5] Unless the adversary increased the sequence number artificially which is equivalent to a substitution attack.

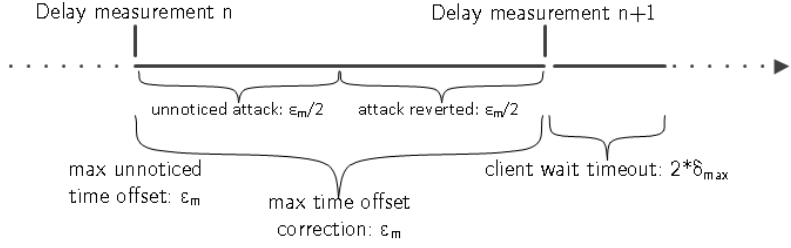

**Fig. 2.** Building blocks of delay bounds.

(2) According to the newly introduced dynamic time offset correction limitation, the maximum time offset correction applied equals $(t_{arr} - t_{last}) \cdot \rho_{max}$. Since the delay measurement interval is fixed (per receiver) and $t_{last}$ is set by the receiver at every delay measurement, the maximum difference $t_{arr} - t_{last}$ equals one delay measurement interval $\frac{\epsilon_m}{\rho_{max}}$. For this reason, the maximum time offset correction that can be applied between two consecutive delay measurements is $\frac{\epsilon_m}{\rho_{max}} \cdot \rho_{max} = \epsilon_m$.

Based on the two building blocks, we now show that the maximum time offset that goes unnoticed equals $\epsilon_1$. Let's assume that the time offset at a delay measurement is $\epsilon_m$ (the greatest value that goes unnoticed, as just shown before). If the adversary wants to get the delay attack to go unnoticed at the subsequent delay measurement, the delay attack must be conducted and its effects reverted within the same delay measurement interval so that the time offset between sender and receiver is again $\leq \epsilon_m$ at the next delay measurement. The time required to conduct the delay attack is equivalent to the time required to revert its effects because of the dynamic time offset correction limitation. If the effects of the adversary's attack exceeds $\epsilon_m$ at the subsequent delay measurement, the receiver will notice (see Figure 2). For this reason, the adversary needs to make sure that additionally delay introduced between the two delay measurements is reverted by the next delay measurement. Therefore, the maximum effect of a delay attack during a delay measurement interval equals $\frac{\epsilon_m}{2} = \delta_{max} - \delta_{min}$, and the upper bound of the time offset that will go unnoticed therefore equals $\epsilon_m + \delta_{max} - \delta_{min} = 3 \cdot (\delta_{max} - \delta_{min}) = \epsilon_1$.

Finally, we can show that the upper bound of the time offset at the point in time when the receiver notices the delay attack equals $\epsilon_2$. We already know that the maximum time offset at a delay measurement equals $\epsilon_m$ and the maximum time offset correction that can be applied during a delay measurement interval equals $\epsilon_m$ as well. At the subsequent delay measurement, the receiver waits for $2 \cdot \delta_{max}$ for the sender's delay reply to the receiver's request. During that time the receiver's clock may drift at most $2 \cdot \delta_{max} \cdot \rho_{max}$. For this reason, the upper bound of the time offset at the point in time when the receiver notices the delay attack equals $\epsilon_m + \epsilon_m + 2 \cdot \delta_{max} \cdot \rho_{max} = \epsilon_2$.

## 6 Computational and Communication Efficiency

Computational efficiency is of crucial importance for high-precision time synchronization, and communication efficiency is generally important as bandwidth is a worthy resource. To evaluate our approach on data origin authentication, we test the computational and communication efficiency of two different, high-performance digital signature schemes in the context of time synchronization (both with a conjectured security of $128$ bit): Ed25519 [12], an elliptic-curve signature scheme *"carefully engineered at several levels of design and implementation to achieve very high speed without compromising security"* [12], and MQQ-SIG [13], a signature scheme based on multivariate-quadratic quasigroups. Both schemes are designed to provide extremely fast signing and verification operations.

### 6.1 Measurement Setup

To test whether our proposed solution actually delivers the expected performance in practice we conducted experimental measurements using the measurement setup depicted in Figure 3. We set up a stratum 1 NTP server that is synchronized to Coordinated Universal Time (UTC) through a dedicated Global Positioning System (GPS) Pulse per Second (PPS) hardware clock. The server broadcasts time synchronization messages to the local network every 8 seconds, which is the shortest interval possible without making changes to NTP's source code.

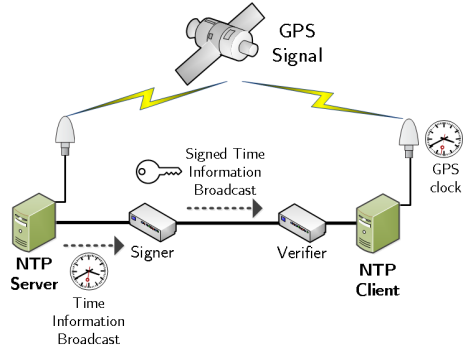

**Fig. 3.** Measurement setup.

On the other end runs an NTP client that synchronizes its clock to the NTP server's broadcast time. To measure the time synchronization delay, the NTP client can access the same GPS PPS timing signal as the server, but is configured to synchronize just to the NTP server and to explicitly ignore the GPS clock in its NTP synchronization algorithm. I.e., the client just logs the accurate GPS clock values to support later assessment of the influence that data origin authentication has onto time synchronization's precision. Both end systems, NTP server and NTP client, use identical time synchronization hardware for synchronization and timestamping. Between the server and the client are two network bridges named Signer and Verifier. Signer waits for NTP time synchronization packets and signs the payload using a high-performance digital signature scheme. Verifier looks for signed NTP time synchronization packets, verifies the signature, and removes it thereafter. At the client, normal, unmodified NTP messages arrive. The described setup allows running unmodified NTP code. Admittedly, signing messages and verifying signatures takes time so that messages arrive delayed to client. The GPS synchronized reference clock at the client allows to quantify this delay and enables measuring the influence on the client's clock synchronization algorithm.

### 6.2 Measurement Results

The main goal of the measurements was to quantify the amount of delay that is introduced to the clock synchronization algorithm when signing and verifying broadcast NTP time synchronization messages. We assumed the introduced delay to be very small due to the use of high-performance signature schemes. To separate the delay introduced by our measurement setup from the delay introduced by signing and verifying messages, we conducted four measurements. Each measurement was run for more than 24 hours.

First, we configured Signer and Verifier as plain network bridges that only forwarded packets. In this scenario, we observed a median delay of 241 µs. This is the influence that Ethernet plus two bridges have on NTP's precision (when the network delay is not compensated). Then, Signer and Verifier not only forwarded packets but first copied the packets to user-space, where they were parsed (but not signed!). This introduced a median delay of 336 µs, which includes the delay for Ethernet as measured before. We, therefore, conclude that copying the packets to user-space and parsing them (two times) introduces around 94 µs of additional delay. The last two measurements finally quantify the amount of delay that is introduced in NTP's clock synchronization algorithm by signing and verifying each packet individually using the signature schemes in question: Ed25519 and MQQ-SIG. For Ed25519, we observed a median delay of 411 µs, which means that about 75 µs were spent on signing messages and verifying signatures. For MQQ-SIG, we observed a median delay of 364 µs. This means that only about 28 µs were spent on signing messages and verifying signatures. Table 2 summarizes the measurement results. We argue that, when implemented within a time synchronization protocol, the delay introduced by the authentication scheme can be even further reduced by (a) the sender communicating the minimum time required to sign a message, and (b) the client subtracting the time it took to verify the signature of a specific message.

**Table 2.** Summary of measurement results

| Measurement | Median Time Offset | Standard Deviation | Relative Time Offset |
|---|---|---|---|
| Ethernet | 241 µs | 5.5 µs | −94 µs |
| Ethernet + nfqueue | 336 µs | 8.7 µs | 0 µs |
| Ethernet + nfqueue + Ed25519 | 411 µs | 8.2 µs | 75 µs |
| Ethernet + nfqueue + MQQ-SIG | 364 µs | 8.2 µs | 28 µs |

The communication overhead is very small because of the signature sizes of both schemes. Again, MQQ-SIG performs better in this regard than Ed25519 (32 B vs. 64 B). Such small signature sizes mean that the resulting communication overhead of the authentication scheme is only 32 B or 64 B per message. For the overall overhead, the size of the sequence number as well as the sizes of the public keys have to be included. The size of the public key is significantly larger for MQQ-SIG than for Ed25519 (32 kB vs. 517 B).

# 7 Security Measures Specific to 2-Step Mode

Most of the previously introduced security measures for the 1-step mode can directly be applied in 2-step mode. But in 2-step mode, there are two messages for time offset correction, SYNC and FOLLOWUP. To facilitate hardware-timestamping and transparent clocks as well as to reduce the impact on precision even further, we sign only the FOLLOWUP message. The sequence number, however, can only ensure that the authentication information in the FOLLOWUP message is fresh – it cannot guarantee that SYNC and FOLLOWUP messages are related in any way. Since we only sign the FOLLOWUP message but not the SYNC message, an adversary could create fake SYNC messages as the sequence numbers are predictable. Such fake SYNC messages injected at malicious points in time would make a receiver adhere to false time. To prevent that, we link the unsigned SYNC message to the corresponding, signed FOLLOWUP message. To this end, we introduce the following extension: the sender includes a number in each SYNC message and includes a hash in the FOLLOWUP message. The hash is the output of a collision-resistant hash function given the SYNC message concatenated with the sender's session public key as input.

The number entailed in each SYNC message must be unpredictable and it also must not repeat within a session. Randomness provides unpredictability and a large sample space ensures probabilistic insignificance so that non-repetition can be provided[6]. For this reason, the random number must be (significantly) larger than the sequence number. Otherwise, random numbers have to repeat within a session (pigeon hole principle), which would facilitate a simple replay attack. The SYNC message makes the hash implicitly include the number, which is essential for establishing a link between SYNC and FOLLOWUP message. By adding the sender's session public key, the receiver can verify that the SYNC message actually belongs to the current session (and has not been replayed from a previous session) because each session uses a new public key. Otherwise, an adversary could just collect SYNC messages with different numbers across sessions and replay them at a later point in time.

The receiver can immediately discard SYNC messages that have a smaller sequence number than the last seen sequence number. But the receiver must not update its local state until a valid FOLLOWUP messages was received. When the receiver receives a SYNC message that is already in the receiver's buffer, the newly received SYNC message can also be discarded (as it is identical). When receiving a FOLLOWUP message, the receiver first checks the validity of the included signature. Then, the receiver checks whether the the hash of the SYNC message concatenated with the session public key is identical to the hash included in the FOLLOWUP message. The receiver also checks that the sequence number of the SYNC message is exactly one below the FOLLOWUP's sequence number. If this is the case, the SYNC message and also the FOLLOWUP message are accepted as valid (for this session); otherwise, they are discarded. The receiver clears its list of buffered SYNC messages after receiving a (valid) FOLLOWUP message, after a delay measurement, and when a new session is started.

---

[6] Coincidences of the birthday paradox can be taken into account by the counter or the OFB mode of a block cipher [14].

Basically, FOLLOWUP messages, delay measurements, and sessions serve as timeout periods for received SYNC messages.

**Transparent Clocks and Hardware-Timestamping**

As pointed out earlier, transparent clocks and hardware-timestamping are serious security challenges. With the 2-step specific security measures, however, hardware-timestamping of SYNC messages is facilitated as the SYNC messages do not need to be cryptographically signed. Furthermore, we can support transparent clocks which can append the delay they introduced to the original FOLLOWUP message and sign the resulting message themselves. We cannot guarantee, however, that a specific transparent clock is on the path from sender to receiver (an adversary could just have compromised the transparent clock and use its secret key to add artificial delay). Nevertheless, this kind of attack can be seen as a special case of a delay attack which will be analyzed later.

## 8 Security Analysis for 2-Step Mode

In Section 5, we analyzed the security our measures provide against every severe attack in 1-step mode. In this section, we will analyze the security of the 2-step specific security measures.

### 8.1 Substitution attack

We have already shown that the signed message is secure against substitution attacks in 1-step mode and, for this reason, conclude that the signed FOLLOWUP message is equally secure in 2-step mode. We need to show, however, that splitting the time synchronization message into a signed FOLLOWUP and an unsigned SYNC message does not increase the adversary's success probability.

The adversary can substitute parts of the SYNC message and the receiver will still accept the message temporarily, as it is not signed and the receiver cannot verify the authenticity of the message at that point. However, before the arrival time of the SYNC message is used for time offset correction, the receiver waits for the FOLLOWUP message as it contains the send time of the SYNC message. If the hash in the FOLLOWUP message does not match the hash of the SYNC message concatenated with the sender's session public key or if the sequence number of the SYNC message is not exactly one below that of the FOLLOWUP message, the receiver will discard the SYNC message. Now, we distinguish two cases: (1) the new SYNC message that resulted from replacing parts of an old message is identical to a message that the adversary has captured before. This case is equivalent to a replay attack or a delay attack, both of which will be analyzed later. (2) The message is new in a sense that the adversary has not seen it before. Then, the adversary needs the hash of the new, substituted message (concatenated with the session public key) to match the hash in the signed FOLLOWUP message. If the adversary can do that efficiently, however, he can also break the collision resistance of the hash function. If, on the other hand, the adversary decides to wait for a FOLLOWUP message that matches the new random number and drops all other messages meanwhile, the receiver will discard the SYNC message because the sequence number will not match. The probability that the random number in a specific FOLLOWUP message matches equals $\frac{1}{n-l}$ with $n$ the size of the sample space and $l$ the number of random numbers that were already used by the sender in this session. The probability is negligible small when $n$ is chosen sufficiently large. Besides, even if the attack should be successful, the impact on the receiver's notion of time would be very low because of the dynamic offset correction limitation.

### 8.2 Pre-play Attack

In a pre-play attack, the adversary aims to make injected messages appear as if it were sent by the legitimate sender. Since we have already shown that pre-play attacks can be prevented in 1-step mode, which is equivalent to the FOLLOWUP message in 2-step mode, we now focus on SYNC messages.

The adversary can try to guess the random number and inject a SYNC message with the guessed random number and a specific sequence number at an earlier point in time, drop all other SYNC and FOLLOWUP messages in the meantime, and wait for the sender to send the FOLLOWUP message with that sequence number, which is equivalent to a substitution attack described above. In a pre-play attack, the adversary can increase his chances, however, by sending a large number of SYNC messages with different random numbers, since the adversary's attempts do not need to be signed. The success probability is relative to the number of messages the adversary can inject, compared to the size of the sample space (and the number of random numbers already used by the sender in this session). However, the adversary is limited twofold: 1) by the number of messages the adversary can inject and 2) by the maximum number of SYNC messages that the receiver can buffer. If the receiver's buffer is full and it cannot accept any more (valid) SYNC message, it is a DoS attack but not a severe attack in which the receiver's notion of time is maliciously manipulated. Furthermore, even if the adversary succeeds, the maximum impact of the attack is limited by the dynamic time offset correction limitation.

For this reason, the 2-step security mode opens another possibility for an adversary to conduct an attack but such pre-play attack requires more resources and has a lower success probability than a delay attack while it cannot have greater impact on the client's notion of time. For this reason, we are convinced that facilitating hardware-timestamping and transparent clocks while having no impact on the time synchronization's precision outweigh the slightly increased attack surface.

### 8.3 Replay Attack

Again, we focus only on SYNC messages as replayed FOLLOWUP messages are equivalent to the 1-step scenario. In a replay attack, we distinguish two cases: (1) the original message is prevented from reaching the receiver which is equivalent to a delay attack and will be analyzed later, and (2) the receiver receives the original as well as the replayed SYNC message. If the replayed SYNC message is received before the corresponding FOLLOWUP message, then the receiver has already stored an identical

SYNC message and will therefore discard the replayed message. If the replayed message is received after the original FOLLOWUP message, the sequence number of the replayed SYNC message is smaller than the FOLLOWUP's sequence number and the replayed SYNC message will therefore be discarded – unless the adversary waits for a new session when the sequence number fits again. Then, however, the receiver will accept the replayed message only when the hash in the FOLLOWUP message matches (note that the hash also includes the sender's session public key). If an adversary can construct such matching hash efficiently, he can also break the collision-resistance of the hash function. For this reason, replay attacks can prevented entirely - also in 2-step mode.

### 8.4 Delay Attack

We have shown that delay attacks can be bound in 1-step mode by a combination of dynamic time offset limitation, a receiver-specific delay measurement interval, and an improved delay measurement procedure. To evaluate whether an adversary can benefit from delaying SYNC or FOLLOWUP messages in the 2-step mode, we distinguish two scenarios: (1) only the SYNC message is delayed, and (2) only the FOLLOWUP message is delayed.

(1) If the delayed SYNC message arrives after the corresponding FOLLOWUP message, the delayed SYNC message will be discarded by the receiver because the message's sequence number is smaller than the receiver's local state updated from the FOLLOWUP message. If the delayed SYNC message arrives before the corresponding FOLLOWUP message, the receiver will accept the delayed SYNC message as valid and set its time according to the (malicious) delay. This is equivalent to the delay attack in 1-step mode and therefore equally bound.

(2) If the FOLLOWUP message is maliciously delayed and there is another FOLLOWUP message or a delay measurement before the delayed FOLLOWUP message, the delayed message will be discarded by the receiver because its sequence number is too small. Therefore, the adversary can delay the FOLLOWUP message only slightly. But there is no benefit for the adversary by delaying the FOLLOWUP message slightly as the FOLLOWUP message contains only time information but its timing has no influence; delaying the FOLLOWUP message only delays the point in time when the receiver will apply the time offset correction but it cannot influence the time offset correction itself.

## 9  Related Work

PTP includes an experimental security extension, Annex K [2], which provides message integrity and replay protection. Annex K, however, is based on symmetric-key cryptography and, therefore, cannot provide data origin authentication. Furthermore, several flaws were discovered and it was never properly formalized [20, 21]. NTP, on the other hand, incorporates two integrated security mechanisms to provide authenticity and integrity: (1) symmetric [1] and (2) Autokey [22]. Both mechanisms cannot provide data origin authentication. Furthermore, the use of Autokey is strongly discouraged as

Table 3. Comparison of SecureTime to related work

| Name | Data Origin Authentication | NTP Support | PTP Support | Substitution Prevention | Pre-play Prevention | Replay Prevention | Delay Mitigation |
|---|---|---|---|---|---|---|---|
| NTS [15, 16, 17] | ✓ | ✓ | ✗ | ✓ | ✓ | ✓ | ✗ |
| PTP extension [18, 19] | ✓ | ✗ | ✓ | ✓ | ✗ | ✓ | ✗ |
| SecureTime | ✓ | ✓ | ✓ | ✓ | ✓ | ✓ | ✓ |

Supported (✓), or not supported (✗).

severe security weaknesses of the algorithm have been discovered [23, 24]. As potential successors of Autokey another proposal is discussed at the Internet Engineering Task Force (IETF): Network Time Security (NTS) [15, 16, 17]. NTS consist of a set of IETF drafts that aim at providing authenticity and integrity for unicast and broadcast time synchronization protocols. By now, NTS is only specified for NTP. For data origin authentication, NTS employs Timed Efficient Stream Loss-tolerant Authentication (TESLA) [25, 26, 27] which was shown to be susceptible to message delay attacks in the context of time synchronization [9]. While NTS prevents substitution attacks and pre-play attacks, delay attacks were not addressed specifically. Replay protection is realized with TESLA but it requires only one message to be sent per time interval which is rather inefficient and may facilitate a simple DoS attack on the receivers' buffers.

A recently suggested security extension to PTP [18, 19] aims to secure the 2-steps of PTP. It provides data origin authentication by employing Ed25519, prevents substitution attacks, and provides replay protection with sequence numbers. However, since the authors use several threat models and did not conduct a comprehensive security analysis, their sequence number window approach facilitates DoS attacks were an adversary just needs to inject SYNC messages with the highest acceptable sequence number. In this way, no other, valid messages will be accepted by the receiver anymore. Furthermore, their proposal is susceptible to pre-play attacks because sequence numbers are predictable and there is no other link between SYNC and FOLLOWUP messages other than the predictable sequence numbers. In this way, an adversary can make receivers adhere to a false time by pre-playing SYNC messages. Delay attacks are also not considered.

## 10 Conclusion

A security measure integrated into time synchronization protocols is required in order to achieve high-precision time synchronization. First, we conducted a comprehensive threat analysis of time synchronization protocols in the context of multicast communication. To prevent the attacks identified in the threat analysis, we suggest a novel set of security measures to secure multicast time synchronization. At the heart of the security measures is a data origin authentication scheme based on high-performance digital signature schemes (Ed25519 or MQQ-SIG). The security measures, furthermore, entail a set of means specific to time synchronization: sequence number and session keys to prevent replay attacks as well as a novel set of countermeasures to mitigate delay attacks consisting of a secured delay measurement procedure, dynamic offset correction limitation, and a receiver-specific delay measurement interval based on one-way network delays and clock drift. We analyzed the security that our security measures provide against every severe attack; substitution attacks, pre-play attacks, and replay attacks are prevented entirely, while delay attacks are mitigated. For delay attacks, we provide upper bounds on the delays that can be introduced maliciously. Additionally, we conducted experimental measurements that show that the security measures we suggest achieve the desired computational and communication efficiency in 1-step mode. We then extended our security measures to support the 2-step mode by requiring only the FOLLOWUP message to be signed and, in this way, support both hardware-timestamping and transparent clocks. These security measures specific to the 2-step mode have practically no impact on the precision of time synchronization.

We are confident that the security measures we suggest in this paper, provide a significant step forward to secure multicast time synchronization.